\documentclass[fleqn,usenatbib]{mnras}

\usepackage[T1]{fontenc}
\usepackage{ragged2e}
\usepackage{color}
\DeclareRobustCommand{\VAN}[3]{#2}
\let\VANthebibliography\thebibliography
\def\thebibliography{\DeclareRobustCommand{\VAN}[3]{##3}\VANthebibliography}
\newcommand{\cii}{[C\,{\sc ii}]}
\newcommand{\ci}{[C\,{\sc i}]}
\newcommand{\nii}{[N\,{\sc ii}]}
\newcommand{\hii}{H\,{\sc ii}}

\newcommand{\oiii}{[O\,{\sc iii}]}
\newcommand{\oi}{[O\,{\sc i}]}

\newcommand{\oiiitocii}{$L_\mathrm{[O\,\textsc{iii}]}/L_\mathrm{[C\,\textsc{ii}]}$}

\newcommand{\oiiil}{[O\,{\sc iii}] 88\,$\mu{\rm m}$}
\newcommand{\ciil}{[C\,{\sc ii}] 158\,$\mu{\rm m}$}

\definecolor{referee}{RGB}{0,0,0}

\usepackage{graphicx}	
\usepackage{amsmath}	
\usepackage{amssymb}	
\usepackage{newtxtext,newtxmath}

\title[O$^{++}$ in four DSFGs]{Breathless BEARS: [O\,{\sc iii}] 88\,$\mu{\rm m}$ Emission of Dusty Star-Forming Galaxies at $z = 3-4$}

\author[Bakx et al.]{T. J. L. C. Bakx$^{1}$\thanks{E-mail: tom.bakx@chalmers.se},
Hiddo S. B. Algera$^2$,
Prachi Prajapati$^{3,4}$,
George Bendo$^{5}$,
Stefano Berta$^{6}$,\newauthor{}
Laura Bonavera$^{7,8}$,
Pierre Cox$^{9}$,
Joaquin Gonz\'alez-Nuevo$^{7,8}$,
Masato Hagimoto$^{10}$,\newauthor{}
Kevin Harrington$^{11,12,13,14}$,
Matthew Lehnert$^{9,15}$,
Stephen Serjeant$^{16}$,
Pasquale Temi$^{17}$,\newauthor{}
Paul van der Werf$^{18}$, 
Chentao Yang$^{1}$, and
Gianfranco De Zotti$^{19}$
\\
$^{1}$Department of Space, Earth and Environment, Chalmers University of Technology, SE-412 96 Gothenburg, Sweden \\
$^2$Institute of Astronomy and Astrophysics, Academia Sinica, 11F of Astronomy-Mathematics Building, No.1, Sec. 4, Roosevelt Rd, Taipei 106319, Taiwan, R.O.C. \\
$^3$I. Physikalisches Institut, Universit\"at zu K\"oln, Z\"ulpicher Strasse 77, D-50937 K\"oln, Germany \\
$^4$Max-Planck-Institut f\"ur Radioastronomie, Auf dem H\"ugel 69, 53121 Bonn, Germany\\
$^{5}$UK ALMA Regional Centre Node, Jodrell Bank Centre for Astrophysics, Department of Physics and Astronomy, \\University of Manchester, Oxford Road, Manchester M13 9PL, UK\\
$^{6}$Institut de Radioastronomie Millimétrique (IRAM), 300 Rue de la Piscine, 38400 Saint-Martin-d’H\`{e}res, France\\
$^{7}$Departamento de Fisica, Universidad de Oviedo, C. Federico Garcia Lorca 18, E-33007 Oviedo, Spain\\
$^{8}$Instituto Universitario de Ciencias y Tecnologas Espaciales de Asturias (ICTEA), C. Independencia 13, E-33004 Oviedo, Spain\\
$^{9}$Institut d’Astrophysique de Paris, Sorbonne Universit\'{e},UPMC Universit\'{e} Paris 6 and CNRS, UMR 7095, 98 bis boulevard Arago, F-75014 Paris, France\\
$^{10}$Department of Physics, Graduate School of Science, Nagoya University, Aichi 464-8602, Japan \\ 
$^{11}$Joint ALMA Observatory, Alonso de C\'ordova 3107, Vitacura, Casilla 19001, Santiago de Chile, Chile\\
$^{12}$National Astronomical Observatory of Japan, Los Abedules 3085 Oficina 701, Vitacura 763 0414, Santiago, Chile\\
$^{13}$European Southern Observatory, Alonso de C\'ordova 3107, Vitacura, Casilla 19001, Santiago de Chile, Chile\\
$^{14}$Instituto de Estudios Astrof\'isicos, Facultad de Ingenier\'ia y 455 Ciencias, Universidad Diego Portales, Av. Ej\'ercito Libertador 441,\\
Santiago, Chile\\
$^{15}$Universit\'e Lyon 1, ENS de Lyon, Centre de Recherche Astrophysique de Lyon (UMR5574), 69230 Saint-Genis-Laval, France\\
$^{16}$School of Physical Sciences, The Open University, Milton Keynes, MK7 6AA, UK\\
$^{17}$Astrophysics Branch, NASA—Ames Research Center, MS 245-6, Moffett Field, CA 94035, USA\\
$^{18}$Leiden Observatory, Leiden University, PO Box 9513, NL-2300 RA Leiden, Netherlands\\
$^{19}$INAF, Osservatorio Astronomico di Padova, Vicolo Osservatorio 5, 35122 Padova, Italy\\
}

\date{Accepted 2026 January 14. Received 2026  January 14; in original form 2025 December 5}

\pubyear{2026}

\begin{document}
\label{firstpage}
\pagerange{\pageref{firstpage}--\pageref{lastpage}}
\maketitle

\begin{abstract}
We present \oiiil{} observations towards four {\it Herschel}-selected dusty star-forming galaxies (DSFGs; $\log_{10} \mu L_{\rm IR} / $L$_{\odot} = 13.5 - 14$ at $z = 2.9-4$) using the Atacama Compact Array (ACA) in Bands 9 and 10. We detect \oiii{} emission in all four targets at $> 3 \sigma$, finding line luminosity ratios ($L_{\rm [O_{III}]} / L_{\rm IR} = 10^{-4.2}$ to $10^{-3}$) similar to local spiral galaxies, and an order of magnitude lower when compared with local dwarf galaxies as well as high-redshift Lyman-break galaxies. Using the short-wavelength capabilities of the ACA, these observations bridge the populations of galaxies with \oiii{} emission at low redshift from space missions and at high redshift from ground-based studies.
The difference in \oiii{} emission between these DSFGs and other high-redshift galaxies reflects their more evolved stellar populations ($>10$~Myr), larger dust reservoirs ($M_{\rm dust} \sim 10^{9 - 11}$~M$_{\rm \odot}$), metal-rich interstellar medium ($Z \sim 0.5-2 $~Z$_{\odot}$), and likely weaker ionization radiation fields. 
Ancillary \cii{} emission on two targets provide \oiiitocii{} ratios at $0.3 - 0.9$, suggesting that ionized gas represents a smaller fraction of the total gas reservoir in DSFGs, consistent with theoretical models of DSFGs as transitional systems between gas-rich, turbulent disks and more evolved, gas-poor galaxies. Expanding samples of DSFGs with \oiii{} emission will be key to place this heterogeneous, poorly-understood galactic phase in its astrophysical context.
\end{abstract}




\section{Introduction}
Dusty star-forming galaxies (DSFGs) are the most active star-forming galaxies in the Universe \citep{blain2002,Casey2014}. With total infrared luminosities exceeding 10$^{12}$ L$_{\odot}$, DSFGs reach the limit of ``maximum starburst'' with star formation rates of 1000~M$_{\odot}$ yr$^{-1}$ or more \citep[e.g.,][]{neri2020}. Compared to local ultra-luminous infrared galaxies (ULIRGs), DSFGs are at least one order of magnitude more numerous and contribute strongly to the cosmic star-formation history \citep{Madau2014,Zavala2021}. Their exact nature remains debated \citep{narayanan2015,Lovell2021MNRAS.502..772L}, and although many are probably mergers \citep{tacconi2008,Bakx2024ANGELS}, clues to the origins of this galaxy phase can be found through the detailed characterization across multiple phases of the interstellar medium (ISM; \citealt{Carilli2013,Hodge2021}). 

Recent comprehensive studies of the molecular and neutral phases have revealed significant insights into their ISM properties. At high densities and lower ionization states, observations of dust, CO, \ci{}, and \cii{} have shown near-constant dust-to-gas ratios \citep{Hagimoto2023}, strong variations in gas depletion timescales \citep{Berta2023,Bakx2024ANGELS,Prajapati2025}, diverse CO spectral line energy distributions \citep{Dannerbauer2009,Daddi2015A&A...577A..46D,Harrington2021}, and complex morphologies and kinematics \citep{Hodge2019,Rizzo2020,Rizzo2024,Tsukui2024}. These studies demonstrate that DSFGs represent a dusty phase of galaxy evolution but are not a homogeneous population, with orders-of-magnitude differences in gas densities and properties \citep{Bakx2024,Prajapati2025}.
\begin{figure*}
    \centering
    \includegraphics[width=0.85\linewidth]{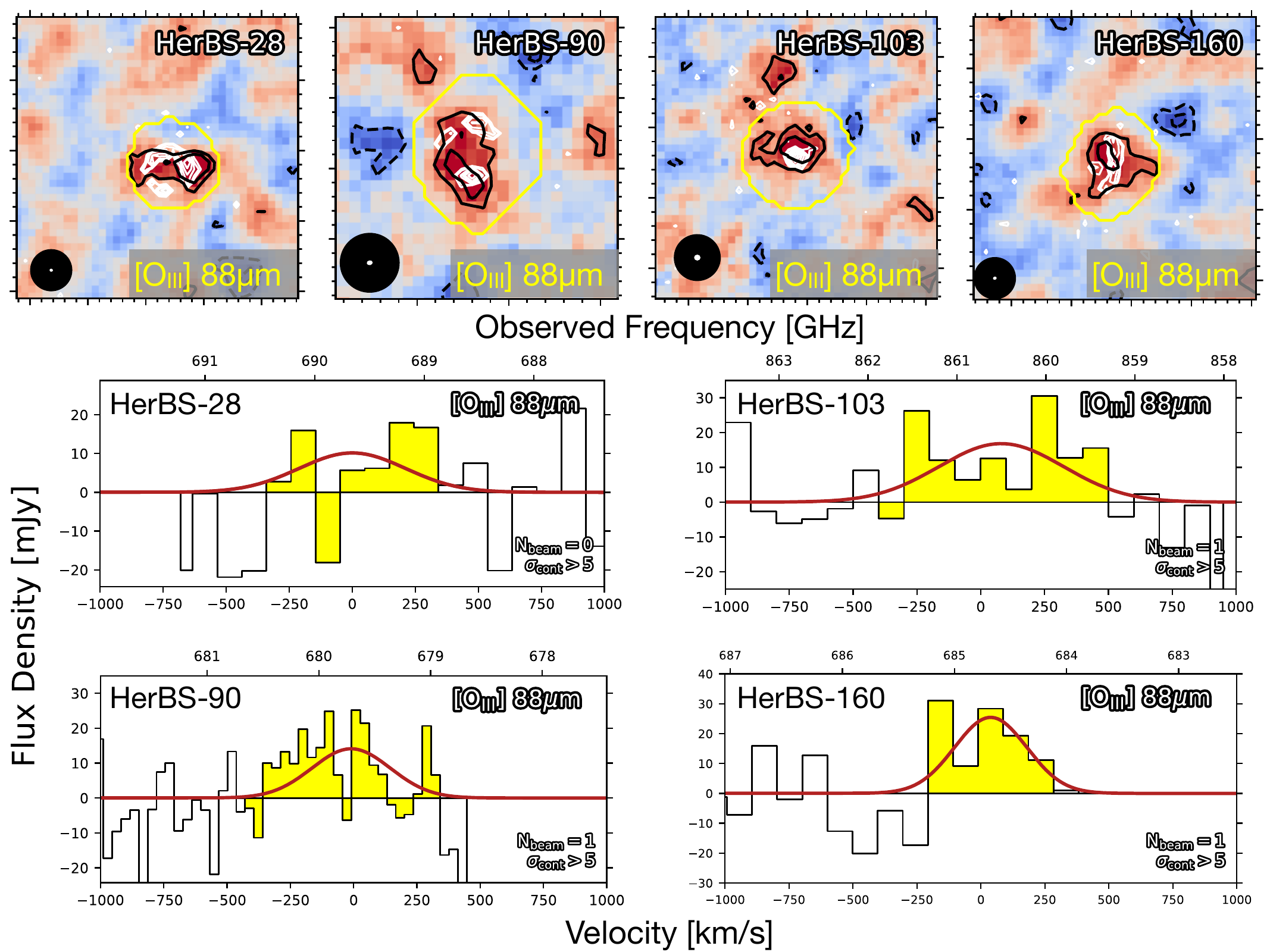}
    \caption{The spectra of the \oiii{} emission line of HerBS-28, HerBS-90, HerBS-103 and HerBS-160. The poststamps shows the 12" by 12" moment-0 map of the \oiiil{} emission line as the background and black solid (positive) and dashed (negative) contours at $2,3,5 \sigma$ starting at $\pm 2$, centered on the positions listed in Table~\ref{tab:sourcesAndLines}. The white contour indicates the high-resolution Band~7 continuum data \citep{Bakx2024ANGELS}, which is used as the basis for the aperture as shown in \textit{yellow contours}. This aperture uses the $5 \sigma$ continuum flux contours, and is subsequently extended by a number of beams (N$_{\rm beam}$) to capture most of the line flux without diluting the signal with excess noise. The spectra show the emission line as a function of velocity (bottom $x$-axis) and frequency (top $x$-axis), where the \textit{filled yellow region} indicates the range over which the moment-0 map is generated, and the \textit{red solid line} indicates a single-Gaussian fit to the line profile.  }
    \label{fig:lineprofiles}
\end{figure*}

Characterizing the varying ISM conditions in this diverse population requires large samples of DSFGs. 
The \textit{Herschel Space Observatory} increased the number of known DSFGs from {\color{referee} thousands (e.g., \citealt{Simpson2019,Smail2021})} to hundreds of thousands through surveys such as H-ATLAS \citep{eales2010,valiante2016,maddox2018,Ward2022} {\color{referee} and {\it Herschel} PEP/HerMES \citep{oliver2012}.} 
Major progress in the past decade includes the development of large DSFG samples with robust redshifts through spectroscopic surveys \citep{reuter20,Cox2023}. The Bright Extragalactic ALMA Redshift Survey (BEARS) exemplifies this approach, conducting ALMA observations of 85 DSFGs to measure spectroscopic redshifts \citep{Urquhart2022}, study field multiplicities \citep{Hodge2013,Karim2013MNRAS.432....2K}, characterize dust spectral energy distributions \citep{Bendo2023}, and model ISM conditions \citep{Hagimoto2023}. These redshift surveys now enable targeted spectral line studies to explore all phases of the ISM systematically.

\begin{table*}
    \centering
    \caption{Sample and line properties of the $z = 3 - 4$ DSFG sample.}
    \label{tab:sourcesAndLines}
    \begin{tabular}{lcccccccccc} \hline
    Source & $z$ & RA & Dec. & $S_{\mathrm{[O\,\textsc{iii}]}}dV$ &  FWHM$_{\mathrm{[O\,\textsc{iii}]}}$ & $L_{\mathrm{[O\,\textsc{iii}]}}$  & $S_{\mathrm{[C\,\textsc{ii}]}}dV$ &  FWHM$_{\mathrm{[C\,\textsc{ii}]}}$ & $L_{\mathrm{[C\,\textsc{ii}]}}$ & \oiiitocii{} \\ 
     & & [hms] & [dms] &  [Jy km/s] & [km/s] & [$10^9$ L$_{\odot}$] &  [Jy km/s] & [km/s] & [$10^9$ L$_{\odot}$] &  \\ \hline
 HerBS-28 & 3.925 & 23:08:15.76 & -34:38:00.3 & $5.6 \pm 1.5$ & \textit{600} & $5.2 \pm 1.5$ & 33.6 $\pm$ 8.2 & 601 $\pm$ 111 & $18.8 \pm 4.2$ & $0.3 \pm 0.1$\\ 
 HerBS-90 & 3.992 & 00:56:59.27 & -29:50:39.7 & 5.7 $\pm$ 3.4 & 364 $\pm$ 163 & $5.4 \pm 3.2$ & - & - & - & - \\
 HerBS-103 & 2.942 &  22:53:24.23 & -32:35:04.3 & 12.3 $\pm$ 2.7 & 576 $\pm$ 128 & $7.1 \pm 1.6$ & 26.0 $\pm$ 7.3 & 669 $\pm$ 144 & $8.5 \pm 2.3$ & $0.9 \pm 0.3$ \\ 
 HerBS-160 & 3.955 &  01:10:14.46 & -31:48:16.2 & 8.3 $\pm$ 3.5 & 297 $\pm$ 96\, & $5.6 \pm 2.2$ & - & - & - & -  \\ \hline
    \end{tabular}
    \raggedright \justify 
\textbf{Notes:} 
Col. 1: Source Name.
Col. 2: Redshift based on CO observations \citep{Urquhart2022}.
Col. 3 \& 4: RA and Dec positions from high-resolution observations.
Col. 5, 6 \& 7: \oiiil{} velocity integrated line fluxes, line widths and luminosities. Limits are shown at $3 \sigma$. Italics indicate a fixed line width based on the \ciil{} width.
Col. 8, 9 \& 10: \ciil{} velocity integrated line fluxes, line widths and luminosities for the two sources where these are available.
Col. 11: The \oiiitocii{} line luminosity ratio.
\end{table*}

However, one crucial component of the ISM remains woefully unexplored: the higher ionization states and densities within these systems \citep{vanderwerf2010,Rybak2022,Harrington2025}. {\color{referee} DSFGs' dusty nature precludes consistent optical follow-up through traditional ionization diagnostics (c.f., \citealt{Shivaei2018}), leaving fundamental questions about their star formation processes unanswered, even with the discerning power of the {\it James Webb Space Telescope} \citep{Hodge2025JWST,Sun2025}.} 
The submillimeter line of doubly-ionized oxygen, \oiiil{}, is emitted close to the center of \hii{} regions and serves as an excellent probe of metallicity \citep{Vallini2015,Olsen2017}, gas density \citep{ferraraCII2019}, and ionization parameter \citep{Harikane2019}. 
Unfortunately, atmospheric transmission precludes ground-based observations of \oiiil{} for galaxies in the local Universe out to Cosmic Noon ($z \sim 2 - 4$; \citealt{Ferkinhoff2010}), and instead current samples of dusty starbursts with \oiii{} observations are dominated by distant systems ($z\gtrsim6$) with incidental follow-up \citep[e.g.,][]{Inoue2016,Marrone2018,Tamura2019,Tadaki2022,Schouws2024,Carniani2025,Algera2024R25,Decarli2025}. {\color{referee} 
Despite Cosmic Noon representing the peak epoch for both cosmic star-formation rate density and AGN activity, seeing to the formation of half of all present-day stars \citep{Madau2014}, benchmarks of the ionization conditions in the most star-forming galaxies during this period remain sparse and incomplete.}

Here we present \oiiil{} observations of four bright southern \textit{Herschel}-selected submillimeter galaxies at $z = 3 - 4$ using the Atacama Compact Array (ACA) or ``Morita Array'', significantly adding to the available data at cosmic noon. This study represents the first systematic survey of \oiii{} emission in gravitationally lensed DSFGs, providing crucial constraints on the ionized medium properties during the peak epoch of cosmic star formation.\footnote{Throughout this letter, we assume a flat $\Lambda$-CDM cosmology with the best-fit parameters derived from the \textit{Planck} results \citep[][paper VI]{Planck2020}, which are $\Omega_\mathrm{m} = 0.315$, $\Omega_\mathrm{\Lambda} = 0.685$ and $h = 0.674$.}

\section{Sample, Observations and Line Fluxes}
We selected our targets from the complete sample of 72 luminosity-limited \textit{Herschel}-detected DSFGs with robust spectroscopic redshifts from the BEARS survey \citep{Urquhart2022}. Initially selected from the H-ATLAS survey \citep{eales2010,valiante2016}, the BEARS sources are all \textit{Herschel} sources in the South Galactic Pole field with $S_{500}$ $>$ 80 mJy and the characteristic red submillimeter colors (250, 350 and 500~$\mu$m; \citealt{pearson13}) consistent with $z > 2$ \citep{bakx18,Bakx2020Erratum}. Our selection was constrained by the requirement that \oiiil{} falls within favorable atmospheric windows for ACA observations. Of the 72 sources, ten galaxies have redshifts between 2.9 and 4.0 that place \oiii{} in optimal atmospheric transmission windows. The bright far-infrared luminosities (log $ \mu L_{\rm IR}$/L$_{\odot}$ = $13.5-14.0$; \citealt{bakx18,Bakx2020Erratum}) derived from \textit{Herschel} fluxes enable the ACA to achieve detections, or deep limits, on the oxygen-to-infrared luminosities within ``reasonable'' ($\sim 8-10$ hour per target) observation times.

The availability of nearby calibrators and the necessity for good weather conditions restricted the final sample with \oiiil{} observations to four DSFGs (HerBS-28, -90, -103, and -160), with one additional source (HerBS-14) having observations with noisy phase calibration data. 
Within specific redshift shells with good atmospheric transparency, this selection provides a relatively unbiased view on the brightest DSFGs.

The ACA observations (2023.1.00750.S; P.I. Tom Bakx) were taken between November 2023 and September 2024 in conditions with precipitable water vapours of 0.15 to 1.01~mm, with the full details in Appendix Table~\ref{tab:observationdetails}. The DSFGs HerBS-28, -90 and -160 were observed using Band 9 \citep{Baryshev2015}, and the DSFG HerBS-103 using Band 10 \citep{Uzawa2013}, with the \oiii{} line centered on one of the spectral windows within one of the sidebands while ensuring the best possible atmospheric transmission in the other sideband. The data cubes were produced using the pipeline \textsc{scriptForPI.py}, and CASA version 6.6.0 \citep{casateam2022} \textsc{tclean} procedure with a natural weighting and the Hogbom deconvolver \citep{Hogbom1974}. The resulting per-bin sensitivity of the data cubes are between 6 and 13~mJy in 35~km/s bins with beam sizes ranging between 0.65 to 1.9~arcseconds, also including a 10 per cent flux uncertainty based on the ALMA Technical Handbook \citep{ALMATECHNICALHANDBOOK2019athb.rept.....R}. 

{\color{referee}
We detect the underlying dust continuum at $> 15 \sigma$ in all four sources. These ACA continuum fluxes are consistent within 20~per cent of the \textit{Herschel} SPIRE 350 and 500 $\mu$m (860 and 600~GHz, resp.) measurements that probe similar observed-frame frequencies, confirming that no significant flux is lost in the ACA observations while revealing no source multiplicity. Since their inclusion in dust continuum modeling does not significantly improve the $L_{\rm IR}$ estimates previously established in \cite{bakx18,Bakx2020Erratum}, we use the established $L_{\rm IR}$ values while continuum-subtracting the data cubes to explore the line emission.}

Line fluxes are extracted using the procedure detailed in \cite{Bakx2024ANGELS}{\color{referee} and summarized in Table~\ref{tab:sourcesAndLines}. We show their continuum-subtracted lines in Figure~\ref{fig:lineprofiles}}. In short, an aperture is created that best matches the expected emission based on ancillary data. In this case, we use higher resolution 0.1~arcsec Band~7 data from the ANGELS program \citep{Bakx2024ANGELS} and Bakx et al. in prep where the dust continuum of the sources is strongly detected ($> 20 \sigma$). This continuum data is then smoothed to the resolution of the \oiii{} data cube. Using the average beam of the data cube, the aperture is widened by a beam if the signal-to-noise ratio permitted\footnote{Indicated by $N_{\rm beam}$ in Figure~\ref{fig:lineprofiles}; the extent of subsequent smoothing is decided on a per-source basis, ensuring no significant flux is missed but also minimizing the additional noise that comes from an aperture that is too large.} to ensure extended flux is also included in the flux measurement. 

The line-integrated (i.e., moment-0) maps show $> 3 \sigma$ \oiiil{} emission for all four sources. Since the redshifts are determined robustly from multiple CO lines, this provides confidence in the veracity of these lines. The spectra for HerBS-28 and -103 were noisy, so a lower-resolution data cube was produced with 100~km/s bins. For all but HerBS-28, a Gaussian line fit subsequently measures the velocity-integrated line flux and line width. For HerBS-28, the line velocity width was fixed to the line width from ancillary \ciil{} measurements in order to constrain the fit. The line luminosities are calculated using the equations given in \cite{solomon2005}. 
Two of the sources, HerBS-28 and -103, have ancillary \cii{} observations (2021.1.00265.S; P.I. Dominik Riechers). These data have been processed in the same fashion as the \oiii{} observations, and are also detailed in Table~\ref{tab:sourcesAndLines} and Figure~\ref{fig:ciispectra}.

\section{Implications}

\begin{figure}
\centering
\includegraphics[width=\linewidth]{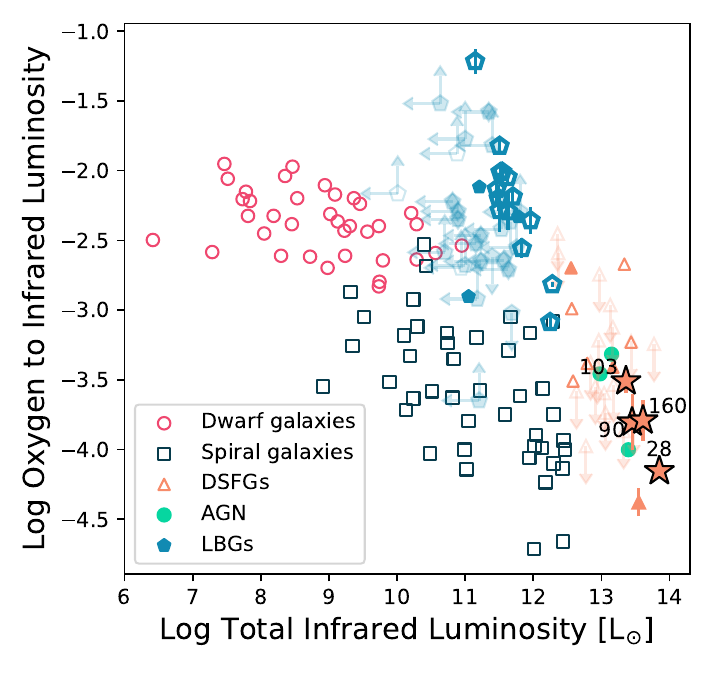}
\caption{The \oiii{}-to-IR luminosity ratio as a function of IR luminosity. The open circles (dwarf galaxies; \citealt{Madden2013,Cormier2015,Cigan2016,Ura2023}) and squares (local spiral galaxies; \citealt{Ferkinhoff2010,HerreraCamus2018}) represent low-redshift galaxies. High-redshift galaxies are shown in open \citep{Bakx2024GoldenRatio} and filled pentagons for Lyman-Break Galaxies \citep{Inoue2016,carniani:2017oiii,Laporte2017, Hashimoto2018,Hashimoto2019,Zavala2024,Bakx2023glass,Schouws2024,Carniani2025,Algera2024,Ren2023,Akins2022,Schouws2021,Fujimoto2024}, with AGN shown in filled circles \citep{Hashimoto2019QSO,Decarli2023} and distant DSFGs in open \citep{Zhang2018} and filled triangles \citep{Marrone2018,Zavala2018DSFG,Tadaki2022}. }
\label{fig:oiiivslfir}
\end{figure}

\begin{figure*}
\centering
\includegraphics[width=\linewidth]{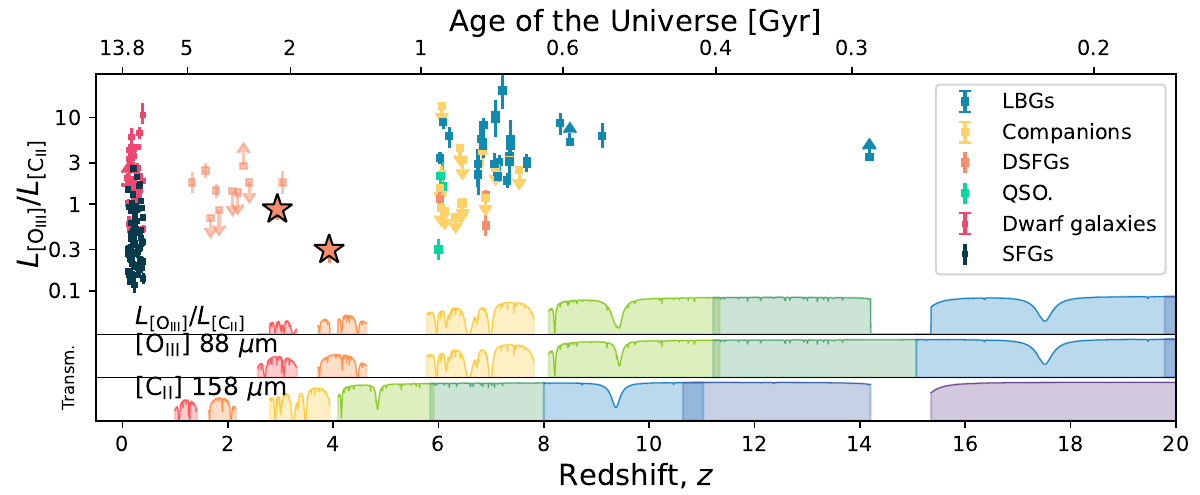}
\caption{Oxygen-to-carbon line ratio (\oiiitocii{}) as a function of redshift from $z = 0$ to 15, showing the observational accessibility of these key diagnostic lines with ALMA. The upper panel displays the \oiiitocii{} ratios measured in various galaxy populations across cosmic time, with our two DSFG sources with both \oiii{} and \cii{} data (\textit{orange stars}) showing low values compared to local dwarf galaxies \citep{Madden2013,Cormier2015,Cigan2016,Ura2023}, and in line with local spiral galaxies \citep{DiazSantos2017} and \textit{Herschel}/FTS-observed galaxies \citep{Zhang2018}. The \oiiitocii{} ratios of high-redshift LBGs \citep{Carniani2025,Schouws2024,Schouws2025,Zavala2024,Zavala2025Halpha,Hashimoto2018,Laporte2019,Fujimoto2024,Tamura2019,Bakx2020,Fudamoto2021Natur,Algera2024,Algera2025OIIIratio,Rowland2024,Inoue2016,Ren2023,Hashimoto2019,Sugahara2022,Akins2022,Wong2022,Carniani2017,Witstok2022,Harikane2019}, companion galaxies \citep{Venemans2020,Bakx2024GoldenRatio,Algera2025OIIIratio},
high-$z$ DSFGs \citep{Marrone2018,Zavala2018DSFG,Tadaki2022} and quasar host galaxies \citep{Hashimoto2019QSO,Walter2018}.
The lower panel shows the atmospheric transmission for \oiiil{} and \cii{} across ALMA’s Bands, with the combined transmission indicating optimal redshift windows for simultaneous observations. 
The systematic \oiii{} deficit in DSFGs is observable across the full redshift range where both lines can be detected, suggesting this is a fundamental property of dusty star-forming environments rather than an observational bias.}
\label{fig:oiiiciivsz}
\end{figure*}

Our ACA observations resulted in four \oiiil{} detections with velocity-integrated line fluxes between 5.6 and 13~Jy~km~s$^{-1}$ and line luminosities ranging from 5 to $10\, \times{}\, 10^{9}$~L$_{\odot}$. Figure~\ref{fig:oiiivslfir} shows the \oiii{}-to-infrared luminosity ratio of our DSFGs against their infrared luminosity, comparing these to the established \oiii{}-infrared luminosity ratios from local dwarf irregular galaxies and spiral galaxies. Our detected DSFGs systematically fall in the regime of star-forming spiral galaxies, with \oiii{}-to-infrared luminosity ratios ranging from $10^{-4.2}$ to $10^{-3}$. 
Lyman-break galaxies (LBGs) at higher redshifts typically show $L_\mathrm{[O\,\textsc{iii}]}$/$L_{\rm IR}$ ratios consistent with or slightly elevated above local dwarf galaxies, while these $z > 2$ DSFGs appear to be \oiii{}-faint for $z > 1$ galaxies.

Figure~\ref{fig:oiiiciivsz} shows the \oiiitocii{} ratio as a function of redshift. Using \cii{} measurements from ancillary ALMA observations, the \oiiitocii{} ratio of HerBS-28 is $0.3 \pm 0.1$ and of HerBS-103 is $0.9 \pm 0.3$. These ratios are significantly lower than those typically observed in high-redshift LBGs ($1-10$; e.g., \citealt[][]{Algera2024}), instead suggesting weaker ionization parameters ($U \sim 10^{-3}$ to $10^{-2.5}$) compared to typical Lyman-break galaxies ($U \sim 10^{-2}$ to $10^{-1.5}$; \citealt{Harikane2019}). {\color{referee}
In these DSFGs with high infrared surface brightnesses \citep{Hagimoto2023,Bakx2024ANGELS}, the fraction of \cii{} emission arising from neutral photo-dissociation regions typically exceeds 80~per cent (e.g., \citealt{DiazSantos2017}). In such environments, the \cii{} line is dominated by the neutral gas reservoir rather than the ionized \hii{} regions from where \oiii{} originates, naturally resulting in the lower observed line ratios compared to less-obscured, high-redshift Lyman-break galaxies.}
This indicates that the ionized gas phase represents a smaller fraction of the total gas reservoir in DSFGs compared to typical star-forming galaxies. This is consistent with the large reservoirs of molecular gas of DSFGs, and in line with recent studies of \cii{}-selected companion galaxies to quasars \citep{Bakx2024GoldenRatio,vanLeeuwen2025}. 


Observationally, it appears that low \oiii{}-to-infrared luminosity ratios are driven by enhanced metallicities \citep{Tamura2019,Hashimoto2019QSO}. Given their large gas masses and high infrared luminosities, DSFGs likely harbor more evolved stellar populations with ages in excess of $>50$~Myr \citep{Berta2023}, while younger LBGs appear to have bursty star-formation \citep{Vallini2021} and younger stellar populations ($\sim 4$~Myr; e.g., \citealt{Tamura2019}). Furthermore, the high dust-to-gas mass ratios of DSFGs appear to indicate (near-)solar metallicities ($Z \sim  0.5-2 $~Z$_{\odot}$; \citealt{Hagimoto2023,Prajapati2025}), while typically LBGs have more modest metallicities (\citealt{Nakajima2023}; c.f., \citealt{Rowland2025}) and dust masses \citep{Inami2022}, in part due to their UV-selection. This interpretation is in line with the elevated \oiiitocii{} ratios of $z\sim 6-8$ observed in high-redshift LBGs, which have been attributed to their bursty star formation histories and young stellar ages \citep{Algera2025OIIIratio}. In contrast, DSFGs likely experience less bursty star formation due to their longer depletion timescales \citep{Hagimoto2023} and more evolved stellar populations \citep{daCunha2015,Berta2023}, naturally resulting in lower \oiii{} emission relative to their total star formation rates. Moreover, \oiii{} emission can also be emitted because of interactions with ionizing particles and other ionization sources within multi-phase gas reservoirs \citep{Harrington2025}. This could complicate the direct relation between the regions where \oiii{} is emitted and the dust emission, since this likely originates from star-forming and more extended, older stellar populations. 

DSFGs are characterized by exceptionally high dust masses ($M_{\rm dust} \approx 10^{9-10}$~M$_{\rm \odot}$; \citealt{Bendo2023,Bendo2025,Hagimoto2023}) resulting in optical depths that significantly exceed those of typical star-forming galaxies \citep{Algera2024,Algera2024R25}. Since the observed dust masses of these HerBS sources approach $10^{11}$~M$_{\odot}$, even accounting for lensing, these sources are likely optically thick until 200~$\mu$m \citep{Casey2018,Tsukui2023}, and selective dust attenuation of \oiii{} could cause the lower ratios. Differential lensing, especially given the high magnification factors of these sources ($\mu > 4$; \citealt{Urquhart2022}) could also affect the line ratios \citep{Serjeant2012}. Appendix Figure~\ref{fig:AttenuatedCurves} quantifies the effect of dust on lines and line ratios, showing that although optical depth can reduce the line-to-infrared ratio, the effect is up to $\sim 0.8$~dex for a mixed medium. This would still place these DSFGs on the low side of the \oiii{}-to-infrared correlation, and suggests that the \oiii{} deficit is not simply a redshift evolution effect but reflects fundamental differences in the ionization structure of dusty versus UV-selected galaxies \citep[c.f.][]{Peng2025}, with less dust-obscured galaxies likely having larger \oiii{}-emitting volumes surrounding star-forming regions. We note that this interpretation can be complicated by geometric effects. The high optical depths and irregular morphologies of DSFGs \citep{Bakx2024ANGELS,Hodge2025JWST} can create significant line-of-sight variations in extinction between lines, exacerbated by the stratified structure of photo-dissociation regions. \oiii{} emission from the innermost, most highly ionized zones could experience preferential extinction compared to \cii{} and the bulk of the infrared emission from the outer, less extinguished regions \citep{Vallini2015}. This effect is seen in observations \citep{Cormier2019,Hagimoto2025} and simulations \citep{narayanan2015,Lovell2021MNRAS.502..772L} of high-redshift LBGs and local dwarf galaxies, with patchy, porous distributions of the neutral ISM, but hard to quantify without extensive, resolved observations.

We can rule out thermal saturation as the primary explanation for our observed line luminosities \citep[c.f.][]{Rybak2019}, as this would preferentially boost \oiii{} as more O$^{++}$ ions become available.  Similarly, an ISM substantially denser than the critical density ($n_{\rm crit} \gg 500$~cm$^{-3}$) could also result in similar results \citep[c.f.][]{Peng2025FLAMESI,Peng2025FLAMESII,Harikane2025,Usui2025}, but this would also affect the \cii{} emission, with critical densities between $n_{\rm crit} = 50$~cm$^{-3}$ in ionized regions and $n_{\rm crit} = 2800$~cm$^{-3}$ in neutral regions \citep{Stacey2011}. This agrees with gas densities estimated for HerBS-90 and -160 based on \ci{}-to-infrared luminosity ratios \citep{Hagimoto2023}, indicating about 0.5~dex denser ISM for HerBS-160 ($10^{4.6}$/cm$^3$ compared to $10^{4.1}$/cm$^3$) but no significant variation in the \oiii{}-to-infrared luminosity ratio.


The implications for our understanding of DSFGs are significant. The \oiii{} deficit relative to other high-$z$ galaxies suggests that, despite their extreme infrared luminosities, DSFGs are not the most extreme ionization environments in the high-redshift Universe. This supports models in which DSFGs represent a phase of galaxy evolution characterized by efficient star formation in gas-rich, metal-rich, and dusty environments. Consequently, the difference in stellar ages and star-formation modes between DSFGs and
UV-selected galaxies contributes to the observed variation in ionization conditions.
Future observations with higher sensitivity and angular resolution will be crucial for testing these interpretations, in particular using the upgraded ALMA capabilities in the Wideband Sensitivity Upgrade \citep{Carpenter2023} and the future space mission PRIMA \citep{Glenn2025}. Spatially resolved \oiii{} maps could reveal whether the deficit reflects global properties or results from complex spatial distributions of ionizing sources and dust. Multi-line diagnostics, including \nii{} and \oi{}, can provide additional constraints on the ionization parameter, electron density, and chemical abundance patterns in these extreme environments.

\section*{Acknowledgements}
{\color{referee} The authors kindly thank Roberto Decarli for his insightful comments and suggestions to improve this manuscript.}
TB gratefully acknowledges financial support from the Knut and Alice Wallenberg foundation through grant no. KAW 2020.0081. MH is supported by Japan Society for the Promotion of Science (JSPS) KAKENHI Grant No. 22H04939. LB and JGN acknowledge the CNS2022-135748 project and the PID2021-125630NB-I00 project. PP is a member of the International Max Planck Research School (IMPRS*) in Astronomy and Astrophysics. PP, DR, and AW acknowledge the Collaborative Research Center 1601 (SFB 1601 sub-project C2) funded by the Deutsche Forschungs Gemeinschaft (DFG, German Research Foundation) – 500700252. HSBA acknowledges support from Academia Sinica through grant AS-PD-1141-M01-2.
We are grateful to Dominik Riechers for the useful discussions. 
This paper makes use of the following ALMA data: ADS/JAO.ALMA 
2021.1.00265.S,
2022.1.00145.S,
2022.1.00432.S, and
2023.1.00750.S. 

\section*{Data Availability}
The datasets are available on the ALMA Science Archive, and generated using the standard data calibration protocols.



\bibliographystyle{mnras}
\bibliography{example} 




\appendix

\section{ACA observation set-ups}
Table~\ref{tab:observationdetails} provides the details of the ACA observations.

\begin{table}
    \centering
    \caption{Parameters of the ACA observations}
    \label{tab:observationdetails}
    \begin{tabular}{lccccccc} 
    \hline
UTC end time          & Frequency   & T$_{\rm int}$ & PWV \\
  & [GHz] 			 & 			 	   [min] &  [mm] \\ \hline
  \multicolumn{4}{c}{\bf HerBS-28} \\\hline
2024-08-10 07:26:44 & 688.930 & 124.16 & 0.47 \\
2024-07-13 10:42:43 & 688.930 & 123.83 & 0.15 \\
2024-06-06 12:27:26 & 688.930 & 123.47 & 0.70 \\
 \hline
  \multicolumn{4}{c}{\bf HerBS-90} \\
 \hline
2024-09-09 08:58:49 & 679.670 & 118.86 & 0.49 \\
2024-09-08 11:03:45	& 679.670 & 122.87 & 0.28 \\
2024-09-08 09:00:40 & 679.670 & 118.66 & 0.30 \\ \hline
  \multicolumn{4}{c}{\bf HerBS-103} \\
 \hline
2024-09-27 07:57:49	& 860.732 & 123.30 & 0.25	 \\
2024-09-27 05:54:26	& 860.732 & 123.34 & 0.24 \\
2024-09-22 06:25:41	& 860.732 & 122.65 & 0.48 \\
2024-09-21 06:14:25	& 860.732 & 122.73 & 0.58	 \\ \hline
  \multicolumn{4}{c}{\bf HerBS-160} \\
 \hline
 2024-07-04 13:39:02	 & 684.764 & 122.55 & 0.47  \\
2024-06-07 13:15:31& 684.764 	& 123.03	& 1.01   \\
2023-11-03 06:09:25& 684.764 	& 123.48	& 0.39   \\
\hline
    \end{tabular}
    \raggedright \justify 
\textbf{Notes:} 
Col. 1: UTC end time of the observations as $[$YYYY-MM-DD hh:mm:ss$]$.
Col. 2: The observed frequency.
Col. 3: The total observation time including overheads.
Col. 4: The precipitable water vapor during the observations.
\end{table}

\section{Ionized Carbon emission of HerBS-28 and -103}
\begin{figure}
    \centering
    \includegraphics[width=\linewidth]{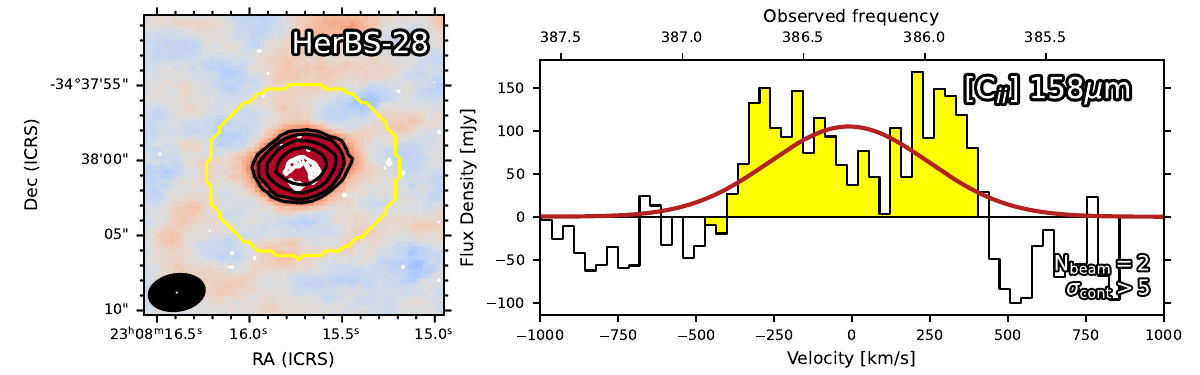}
    \includegraphics[width=\linewidth]{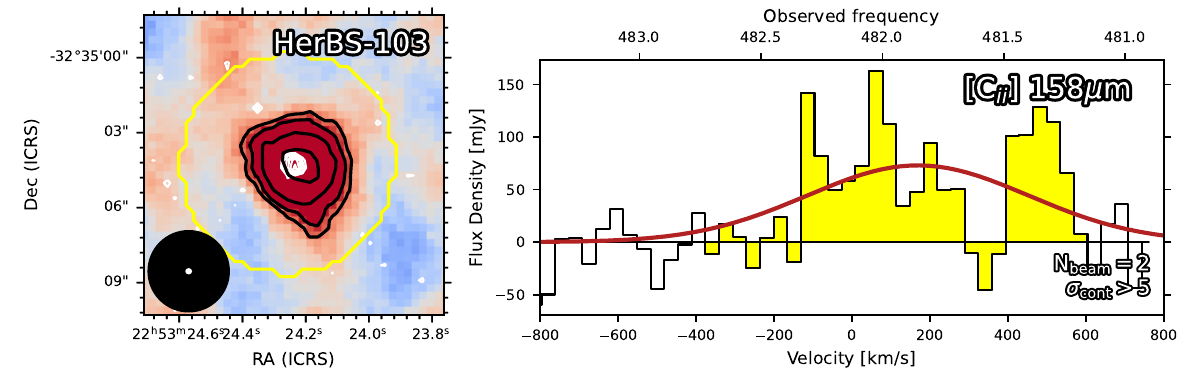}
    \caption{\cii{} spectra of HerBS-28 and HerBS-103, similar to Figure~\ref{fig:lineprofiles}. The poststamps shows the 12" by 12" moment-0 map of the \ciil{} emission line as the background and black contours at $2,3,5 \sigma$ starting at $\pm 2$, centered on the positions listed in Table~\ref{tab:sourcesAndLines}. The white contour indicates the high-resolution Band~7 continuum data \citep{Bakx2024ANGELS}, which is used as the basis for the aperture as shown in \textit{yellow contours} for the spectrum shown on the right-hand side.}
    \label{fig:ciispectra}
\end{figure}

Figure~\ref{fig:ciispectra} shows the \cii{} spectra of HerBS-28 and -103, produced through the same process as the \oiii{} spectra discussed in this paper.

\section{Dust attenuation of line ratios}
Following the prescriptions as detailed in \cite{Decarli2023}, \cite{Algera2024} and \cite{Tadaki2025}, we estimate the effect of dust attenuation on line fluxes and ratios as follows. The optical depth $\tau_{\lambda}$ at a given wavelength $\lambda$ relative to the transitional wavelength $\lambda_{\rm thick}$ is $\tau_{\lambda} = (\lambda / \lambda_{\rm thick})^{- \beta}$, where $\beta$ is the dust emissivity index assumed to be $\sim 2$ in line with most observations of DSFGs \citep{Bendo2023,Bendo2025}. The subsequent transmission is calculated as $F_{\rm obs.} / F_{\rm intr.} = (1 - e^{-\tau_\lambda})/\tau_\lambda$. 

\begin{figure}
    \centering
    \includegraphics[width=\linewidth]{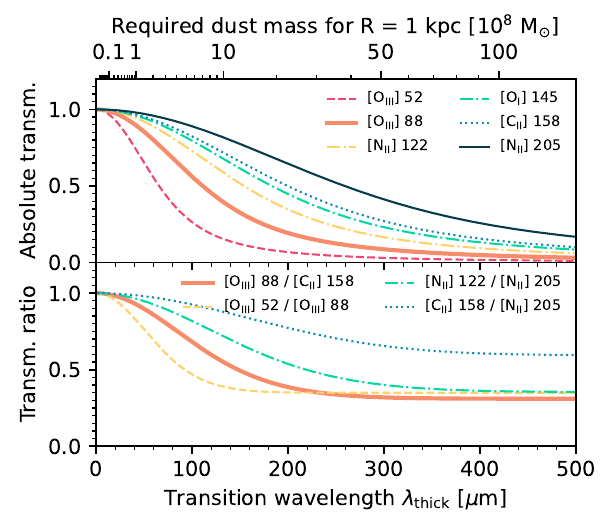}
    \caption{The dust attenuation affects line ratios differently for a varying transitional absorption wavelength, $\lambda_{\rm thick}$. The ratio varies between 1 (no effect of dust attenuation) to $(\lambda_1 / \lambda_2)^\beta$ in the optically-thick case. Assuming Milky Way dust properties, a spherical dust distribution and $\beta_{\rm dust} = 2$, the required dust mass to achieve a transition wavelength is shown in the top x-axis. }
    \label{fig:AttenuatedCurves}
\end{figure}

Figure~\ref{fig:AttenuatedCurves} shows the absolute transmission of lines and relative attenuation of line ratios at different frequencies as a function of the transition wavelength. The absolute transmission of \oiii{} drops from 1 to 0.19 at $\lambda_{\rm thick} = 200$~$\mu$m. The ratio varies between 1 (no effect of dust attenuation) to $(\lambda_1 / \lambda_2)^\beta$ in the optically-thick case. For \oiiitocii{}, this means a maximum dust effect of 31 per cent on the line ratio, with 69 per cent preferential absorption in the \oiiil{} relative to the \ciil{}. Using equation 6 from \cite{Algera2024}\footnote{$\lambda_{\rm thick} \approx 42 {\rm \mu m} \times{} (R\, {\rm [kpc]})^{-1} \times{} (M_{\rm dust}\,[10^8 {\rm M_{\odot}}])^{1/2}$}, we scale the transition wavelength to a total required dust mass, presuming Milky Way dust mass absorption coefficient and a typical dust reservoir size of 1~kpc. 

\bsp	
\label{lastpage}
\end{document}